\documentclass{article}

\usepackage{graphicx}
\usepackage{xcolor}
\usepackage{caption}
\usepackage{cite}
\usepackage{fullpage}
\usepackage{booktabs}
\usepackage{amsmath,amssymb}

\newcommand{\Rey}{\mathrm{Re}}
\newcommand{\SGSsub}{\textsc{sgs}}
\newcommand{\DNSsub}{\textsc{dns}}
\newcommand{\LESsub}{\textsc{les}}

\newcommand{\pp}[2]{\frac{\partial #1}{\partial #2}}

\title{Dynamic Deep Learning LES Closures: Online Optimization With Embedded DNS}

\author{Justin Sirignano\footnote{Mathematical Institute, University of Oxford} \phantom{.} and Jonathan F. MacArt\footnote{Aerospace \& Mechanical Engineering, University of Notre Dame}}

\begin{document}
\maketitle

\begin{abstract}
Deep learning (DL) has recently emerged as a candidate for closure modeling of large-eddy simulation (LES) of turbulent flows. High-fidelity training data is typically limited: it is computationally costly (or even impossible) to numerically generate at high Reynolds numbers, while experimental data is also expensive to produce and might only include sparse/aggregate flow measurements. Thus, only a relatively small number of geometries and physical regimes will realistically be included in any training dataset. Limited data can lead to overfitting and therefore inaccurate predictions for geometries and physical regimes that are different from the training cases. We develop a new online training method for deep learning closure models in LES which seeks to address this challenge. The deep learning closure model is dynamically trained during a large-eddy simulation (LES) calculation using embedded direct numerical simulation (DNS) data. That is, in a small subset of the domain, the flow is computed at DNS resolutions in concert with the LES prediction. The closure model then adjusts its approximation to the unclosed terms using data from the embedded DNS. \emph{Consequently, the closure model is trained on data from the exact geometry/physical regime of the prediction at hand.} An online optimization algorithm is developed to dynamically train the deep learning closure model in the coupled, LES-embedded DNS calculation.   
\end{abstract}

\section{Introduction} \label{Introduction}
Existing machine learning closure models for large-eddy simulation (LES) require high-fidelity data for training \cite{TurbulentWakes,MacArt2021,Sirignano2020}. In virtually all existing cases, the machine learning closure model is trained on high-fidelity direct numerical simulation (DNS) datasets for relatively few different geometries and physical parameters (e.g., Reynolds numbers, Mach numbers, and shape parameters). Then, the machine learning model is simulated out-of-sample for a completely new geometry or physical regime, with no guarantee that the out-of-sample simulation will be accurate. In particular, if the geometry or physical regime is very different from those available in the training set, then the actual fluid flow will likely be evolve differently than the training flow(s). This can lead the deep learning-LES (DL-LES) model to produce inaccurate simulations for the out-of-sample prediction case.  

High-fidelity DNS training data is costly to generate: even for relatively few training cases, DNS can require many times more computational effort than simply increasing the mesh resolution of a classically modeled LES calculation, and DNS for complex geometries at high Reynolds number will be computationally intractable for the foreseeable future. Experimental data can also be expensive to produce and might only include sparse/aggregate flow measurements. Consequently, training datasets are limited and, by definition, can be very different (in terms of geometries, boundary conditions, and physical parameters) than the target prediction cases. Significant differences between the training and test (i.e., prediction) cases can lead to poor accuracy for predictive simulations. 

We develop a new dynamic machine learning closure model that seeks to address these challenges. It couples an LES calculation with embedded DNS calculations, and the LES closure model is dynamically trained on the filtered, embedded DNS data using an online optimization algorithm. Specifically, the majority of the physical domain is modeled by solving the LES equations coupled to one or more deep learning closure models. The flow in a small subset of the domain is evaluated to high accuracy using an embedded DNS, the boundary conditions of which are provided by the LES prediction. The coupled LES and DNS equations are solved simultaneously, and the LES closure model is simultaneously trained using the filtered, embedded DNS data. The advantages of the method are:
\begin{itemize}
\item The closure model is trained on data for the exact geometry and physical conditions for which a prediction is being made. 
\begin{itemize}
\item Embedded DNS data is simultaneously generated during the simulation via coupled LES--embedded DNS equations.
\item Online optimization simultaneously trains the closure model on the filtered embedded DNS data. 
\end{itemize}
\item The training optimizes over the entire LES equations, accounting for both closure and discretization errors in the LES calculation for the target problem.
\item At each simulation step, a stochastic gradient step is taken to update the closure model given the filtered embedded DNS data. The online optimization method will asymptotically optimize over the steady-state statistics of the turbulent flow.
\end{itemize}

We highlight that embedded DNS has been previously developed for closure modeling in the simulation of turbulent flows in recent articles by He and Chen~\cite{LiHeIJNMF,LiHeJFM2022_Paper1,LiHeJFM2022_Paper2}, who pioneered closure methods for RANS and LES using embedded DNS. In their method, DNS is simulated on fine-mesh blocks and then used to estimate a closure model for RANS or space-time-averaged LES using a block-spectral method, without a machine learning model. The major novelty of our work is that (1) we implement a machine learning model in the LES equations that is trained online using the embedded DNS and stochastic gradient descent, and (2) we develop closure methods for LES using only spatial filtering rather than time-averaged LES. In particular, the main contribution of our paper is the development of an online optimization method for the dynamic training of machine learning closure models for coupled embedded DNS--LES simulations.

Section \ref{DynamicMethod} describes the online optimization method for dynamic deep learning closure of LES. Numerical results for decaying isotropic turbulence are presented in Section \ref{Numerical}. Conclusions and future analysis are discussed in Section \ref{Conclusion}.

\section{Dynamic Deep Learning Closure Models for LES} \label{DynamicMethod}

Consider the incompressible Navier--Stokes equations:
\begin{align}
\frac{\partial u_i}{\partial t} &= - \frac{\partial p}{\partial x_i} - \frac{\partial (u_i u_j)}{\partial x_j} + \frac{1}{\mathrm{Re}} \frac{\partial^2 u_i}{\partial x_j^2}, \notag\\
\pp{u_i}{x_i} &= 0,
\label{IncompressibleNS}
\end{align}
where $u = (u_1, u_2, u_3)$. Let $\Omega$ be the domain of the PDE (e.g., $[0, 1] \times [0,1] \times [0,1]$). Define the filtered quantity $\bar{v}(t,x) = \int_{\Omega} G(x - r) v(t, r) dr$, where $G(z)$ is the filter kernel (e.g., a Gaussian or box filter). Applying the filter operation to (\ref{IncompressibleNS}) yields the LES equations for the filtered velocity $\bar{u}$:
\begin{align}
\frac{\partial \bar{u}_i}{\partial t} &= - \frac{\partial \bar{p}}{\partial x_i} - \frac{\partial (\bar{u}_i \bar{u}_j)}{\partial x_j} + \frac{1}{\mathrm{Re} } \frac{\partial^2 \bar{u}_i}{\partial x_j^2} - \frac{\partial}{\partial x_j} (\overline{u_i u_j} - \bar{u}_i \bar{u}_j), \notag \\
\pp{\bar{u}_i}{x_i} &= 0.
\label{IncompressibleLES}
\end{align}
The unclosed term $\tau_{ij}^\SGSsub =  \overline{u_i u_j} - \bar{u}_i \bar{u}_j$ will be modeled with a neural network $h_{ij}(\bar{u}_x; \theta)$, where the parameters $\theta$ must be calibrated to data. In principle, we would like to train the neural network to represent a function relating the LES variables (representing the large turbulent scales) and the subgrid-scale stress tensor $\tau$ (representing the small turbulent scales). If such a closure model $h(\bar{u}_x; \theta)$ can be trained, the LES equation (\ref{IncompressibleLES}) can be simulated at low cost on a coarse mesh. In comparison, simulation of the Navier--Stokes equations (\ref{IncompressibleNS}) requires very small mesh sizes to resolve the smaller turbulent scales, becoming prohibitively expensive at higher Reynolds numbers.  

Previous methods for training machine learning closure models have used off-line optimization. The parameters $\theta$ for the closure model $h(\bar{u}_x; \theta)$ in (\ref{IncompressibleLES}) are trained using high-fidelity data for certain geometries and physical regimes $q_1, \ldots, q_N$. For example, $q_i$ could be a specific flow geometry at different Reynolds numbers. Then, an off-line model is evaluated out-of-sample for a completely new geometry/physical regime $q_{N+1}$ that was not included in the training set. If $q_{N+1}$ is very different than the training set $\{q_1, \ldots, q_N \}$, then the prediction may be inaccurate. 

This paper develops a new dynamic closure model which is trained using online optimization during the actual simulation. In particular, a prediction for a geometry/physical regime $q_{N+1}$ will train a closure model on embedded DNS data from $q_{N+1}$. A coupled LES-embedded DNS calculation will generate its own training data and, at each simulation step, the closure model will be updated with a stochastic gradient descent learning step. 

For the large majority of the physical domain $\Omega$, the LES equations are simulated on a coarse mesh at concomitantly low computational cost. For a small subset of $\Omega$, DNS is simulated with a suitably small mesh size to resolve the Kolmogorov scale. Let $\Omega_\LESsub$ and $\Omega_\DNSsub$ respectively be these two subsets of the domain. The coupled LES--embedded DNS equations are:
\begin{align}
\frac{\partial \bar{u}_i}{\partial t} &= - \frac{\partial \bar{p}}{\partial x_i} - \frac{\partial (\bar{u}_i \bar{u}_j)}{\partial x_j} + \frac{1}{\mathrm{Re} } \frac{\partial^2 \bar{u}_i}{\partial x_j^2} - \frac{\partial h_i}{\partial x_j}(\bar{u}_x; \theta) , \phantom{....} x \in \Omega_\LESsub, \notag \\
\frac{\partial u_i}{\partial t} &= - \frac{\partial p}{\partial x_i} - \frac{\partial (u_i u_j)}{\partial x_j} + \frac{1}{\mathrm{Re} } \frac{\partial^2 u_i}{\partial x_j^2}, \phantom{...} x \in \Omega_\DNSsub,
\label{coupledLESembeddedDNS}
\end{align}
with the additional continuity constraints $\nabla \cdot \bar{u} = 0 $ and $\nabla \cdot u = 0 $. In \eqref{coupledLESembeddedDNS}, $u$ and $\bar u$, as well as their first two derivatives, must of course also match at the interface between $\Omega_\LESsub$ and $\Omega_\DNSsub$. In practice, (\ref{coupledLESembeddedDNS}) is simply the Navier--Stokes equations (\ref{IncompressibleNS}) simulated on a nonuniform mesh where
\begin{itemize}
\item the sub-domain $\Omega_\LESsub$ has a coarse mesh with a closure model $h$, and
\item the sub-domain $\Omega_\DNSsub$ has a fine mesh with no closure model.
\end{itemize}
The velocities $u$ on $\Omega_\DNSsub$ can be filtered to generate target data $\hat{u}$ for training the LES closure model $h$. Specifically, the parameters $\theta$ for the closure model $h(\bar{u}_x; \theta)$ are trained via the online optimization algorithm:
\begin{itemize}
\item At each time step $j$:
\begin{itemize}
\item Simulate $(\bar{u}_j,u_j)$ on $\Omega = \Omega_\LESsub \times \Omega_\DNSsub$ one time step forward to produce $\bar{u}_{j+1}$ on $\Omega_\LESsub$ and $u_{j+1}$ on $\Omega_\DNSsub$.
\item Simulate $\bar{u}_j$ on $\Omega_\DNSsub$ one time step forward, where the one-time-step simulation is initialized from the filtered embedded DNS $\hat{u}_j$, to produce $\bar{u}_{j+1}$ on $\Omega_\DNSsub$.
\item Filter the embedded DNS $u_{j+1}$ to generate the filtered embedded DNS data $\hat{u}_{j+1}$, which will be used for training. 
\item Take a stochastic gradient descent step with learning rate $\alpha^{(j)}$ to update the parameters $\theta^{(j)}$ to minimize the difference between $\bar{u}_{j+1}$ and $\hat{u}_{j+1}$ on the sub-domain $\Omega_\DNSsub$.
\end{itemize}
Note that the parameters $\theta^{(j)}$ are simultaneously optimized as (\ref{coupledLESembeddedDNS}) is simulated, yielding a dynamic closure method for the LES equations. 

\end{itemize}

Our dynamic closure method is specifically designed for predictions of the steady-state statistics of the turbulent flow. Although the flow velocities themselves are unsteady, their time-averaged statistics (e.g., the distribution of the flow velocity) will converge as $t \rightarrow \infty$ \cite{NSergodicity}. Our closure algorithm thus \emph{dynamically} optimizes the closure model parameters $\theta$ to recover the filtered embedded DNS steady-state dynamics as $t \rightarrow \infty$. In principle, as the closure model is improved during the online training, the LES model improves and the boundary conditions of the embedded DNS calculation become more accurate. As the boundary conditions for $\Omega_\DNSsub$ become more accurate, the embedded DNS more accurately represents the flow at the smallest turbulent scales, thus further improving the training of the closure model.

\section{Numerical Results} \label{Numerical}

\subsection{Computational Implementation}

We consider applications to incompressible, decaying, isotropic turbulence. The coupled DNS--LES equations are solved on staggered meshes using second-order central differences in space and the explicit fourth-order Runge--Kutta method in time. The continuity constraint is enforced using a standard pressure-projection method~\cite{Chorin1968}.

We use automatic differentiation to calculate the gradient of the distance between $\bar{u}_{j+1}$ and $\hat{u}_{j+1}$. A constant learning rate of $\alpha^{(j)} = 10^{-1}$ is used for the gradient-descent steps. It is possible that performance can be improved with a learning rate that decreases as a function $j$, which is suggested by the mathematical theory of stochastic gradient descent. 

The DNS calculations are initialized using a synthetic turbulence spectrum~\cite{Passot1987}, which is allowed to decay for approximately one integral time before LES calculations commence. The LES calculations are initialized by filtering the DNS data (using a box filter) and downsampling.\footnote{A collocated filter is used, which introduces some error for the initialization of the staggered mesh LES calculation, and could be improved in the future.} All computations in the simulation are implemented on a graphics processing unit (GPU), including the network closure model, Navier--Stokes solutions, filtering of the embedded DNS, and automatic differentiation.

\subsection{Decaying Isotropic Turbulence}
The dynamic deep learning closure method is implemented on two decaying isotropic turbulence examples with mesh sizes $N=512^3$ and $N=1024^3$. These  are listed in Table~\ref{tab:Re} with their initial turbulent Reynolds numbers $\Rey_{t,0}\equiv \rho u_\mathrm{rms}\ell/\mu$, where $u_\mathrm{rms}=\langle u_i u_i\rangle^{1/2}$  is the domain root-mean-square velocity, $\ell=u_\mathrm{rms}^3/\epsilon$ is the pseudo-integral scale, and $\epsilon$ is the turbulent kinetic energy dissipation rate, and initial Taylor Reynolds numbers $\Rey_{\lambda,0}\equiv \rho u_\mathrm{rms}\lambda/\mu$, where $\lambda$ is the Taylor microscale.
\begin{table}[h]
    \centering
    \begin{tabular}{c c c c}
    \toprule
    Case & $N$ & $\Rey_{\ell,0}$ & $\Rey_{\lambda,0}$ \\
    \midrule
    A & $512^3$  & 1576 & 102.5 \\
    B & $1024^3$ & 1749 & 162.0 \\
    \bottomrule
    \end{tabular}
    \caption{Initial turbulent and Taylor-microscale Reynolds numbers for DNS cases A and B.}
    \label{tab:Re}
\end{table}

In order to evaluate the accuracy of the dynamic deep learning closure method, DNS calculations were separately run on grids of size $1024^3$ and $512^3$ for these two cases. The accuracy of the dynamic DL-LES model is evaluated by comparing against these filtered DNS data.

For both the $1024^3$ and $512^3$ cases, the dynamic DL-LES models the filtered velocities for a box filter. The embedded DNS domain $\Omega_\DNSsub$ is a cube of length $16$ LES computational cells, i.e. $(16 \times d)^3$ embedded DNS computational cells, where $d = \Delta_\LESsub/\Delta_\DNSsub$ is the LES filter size. The number of LES mesh points are $(\frac{512}{d})^3$ and $(\frac{1024}{d})^3$, respectively. For example, in our numerical simulations, the $1024^3$ case with a filtering ratio $d$ will correspond to a LES calculation with $(\frac{1024}{d})^3$ mesh points. For the largest filter sizes ($d=9$ and $d=15$), we use embedded high-fidelity-LES calculations (rather than true DNS) with embedded meshes finer than the full-scale LES but coarser than the DNS, i.e. $\Delta_\LESsub < \Delta_{\textrm{Embedded}} < \Delta_\DNSsub$. For these, the embedded simulation uses no subgrid model. The full-scale DNS, full-scale LES, and embedded mesh sizes are summarized in Table \ref{tab:SimulationGrids}. 

\begin{table}[h]
    \centering
    \begin{tabular}{c c c c c}
    \toprule
     $N_\DNSsub$ & $N_\LESsub$ & $d = \Delta_\LESsub/\Delta_\DNSsub$ & $N_\mathrm{Embedded}$ & $\Delta_\mathrm{Embedded}/\Delta_\DNSsub$ \\
    \midrule
    $512^3$  & $102^3$ & $5$ & $(16 \times 5)^3$ & $1$ \\
    $1024^3$  & $204^3$ & $5$ & $(16 \times 5)^3$ & $1$ \\
    $1024^3$  & $146^3$ & $7$ & $(16 \times 7)^3$ & $1$ \\
    $1024^3$  & $113^3$ & $9$ & $(16 \times 3)^3$ & $3$ \\ 
    $1024^3$  & $68^3$ & $15$ & $(16 \times 5)^3$ & $3$ \\  
    \bottomrule
    \end{tabular}
    \caption{DNS, LES, and embedded simulation grid and filter sizes for test cases.}
    \label{tab:SimulationGrids}
\end{table}

In our numerical results, we compare the dynamic DL-LES model against the filtered DNS. We also compare, as a benchmark, the dynamic DL-LES model against (1) the LES equations with no closure model (i.e., the unclosed terms are set to zero), referred to as ``NM-LES,'' and (2) the LES equations with the constant-coefficient Smagorinsky closure model~\cite{Smagorinsky1963}, referred to as ``SM-LES,'' with $C_S=0.17$. Two versions of the dynamic DL-LES model are trained and evaluated. In the first version, referred to as ``Dynamic DL-LES,'' the closure model is a neural network trained using the online embedded DNS procedure. In the second version, referred to as ``Dynamic DL-SM-LES,'' the closure model is the Smagorinsky model plus a neural network trained using the online embedded DNS procedure.

\subsubsection{Resolved Kinetic Energy Decay}

\begin{figure}[!htb]
  \begin{minipage}{0.48\textwidth}
     \centering
     \includegraphics[width=.9\linewidth]{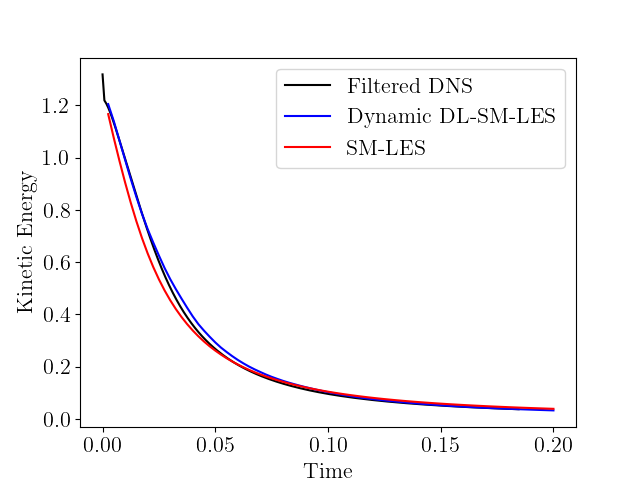}
     \caption{$1024^3$ with $d=7$ filter}\label{Kinetic1024_7x}
   \end{minipage}
      \begin{minipage}{0.48\textwidth}
     \centering
     \includegraphics[width=.9\linewidth]{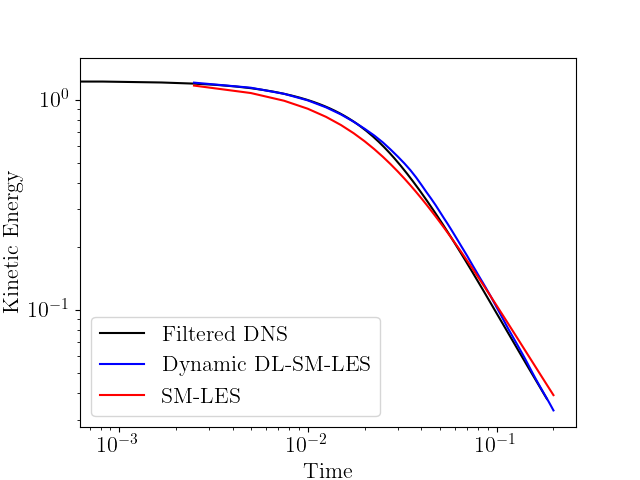}
     \caption{$1024^3$ with $d=7$ filter}\label{Kinetic1024_7x_LogLog}
   \end{minipage}
      \begin{minipage}{0.48\textwidth}
     \centering
     \includegraphics[width=.9\linewidth]{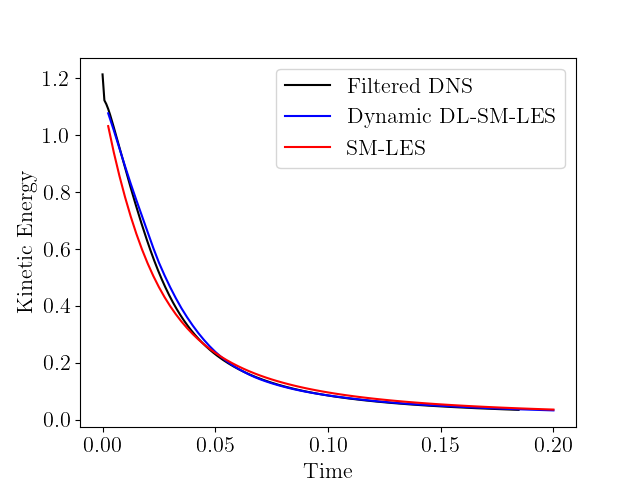}
     \caption{$1024^3$ with $d=9$ filter}\label{Kinetic1024_9x}
   \end{minipage}
         \begin{minipage}{0.48\textwidth}
     \centering
     \includegraphics[width=.9\linewidth]{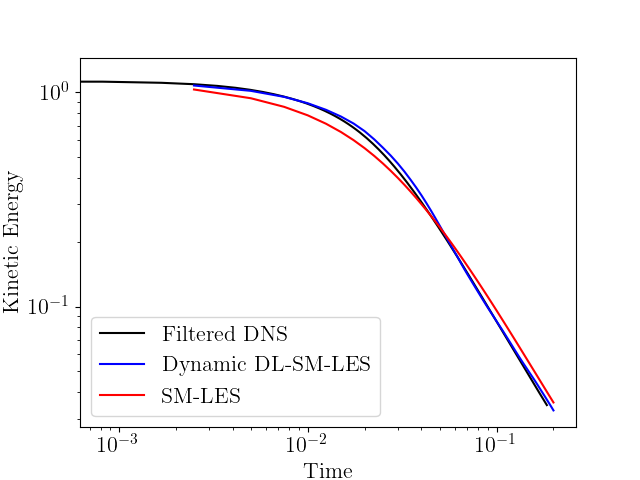}
     \caption{$1024^3$ with $d=9$ filter}\label{Kinetic1024_9x_LogLog}
   \end{minipage}
      \begin{minipage}{0.48\textwidth}
     \centering
     \includegraphics[width=.9\linewidth]{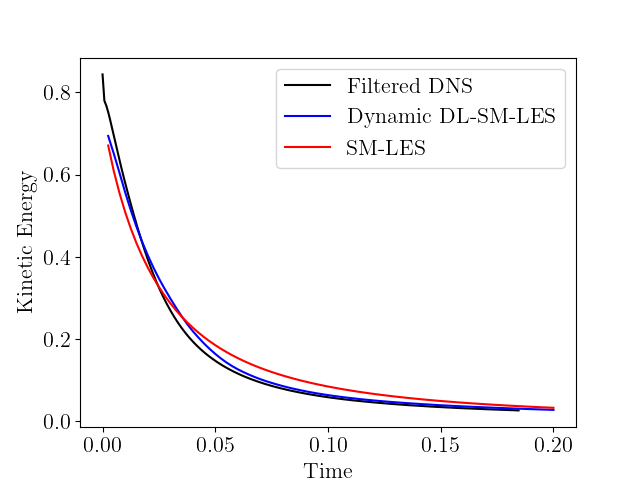}
     \caption{$1024^3$ with $d=15$ filter}\label{Kinetic1024_15x}
   \end{minipage}  
         \begin{minipage}{0.48\textwidth}
     \centering
     \includegraphics[width=.9\linewidth]{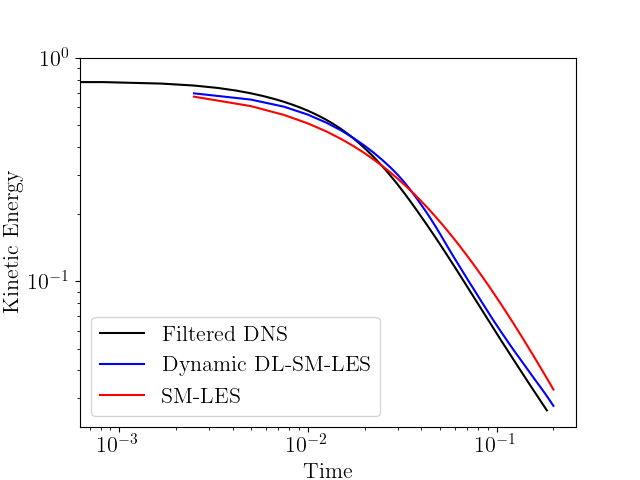}
     \caption{$1024^3$ with $d=15$ filter}\label{Kinetic1024_15x_LogLog}
   \end{minipage}   
    \end{figure}

We compare the resolved kinetic energy decay predicted by the LES models in Figures \ref{Kinetic512}--\ref{Kinetic1024_15x_LogLog}. In these, the plots in the right column are on a log-log scale. 
Figures \ref{Kinetic512}--\ref{Kinetic512_LogLog} compare the kinetic energy of the filtered DNS, dynamic DL-LES, dynamic DL-SM-LES, SM-LES, and NM-LES models for the $N=512^3$ case with $d=5$ filtering. Similarly, Figures \ref{Kinetic1024_5x}--\ref{Kinetic1024_5x_LogLog} compare the LES models for the $N=1024^3$ case with $d=5$ filtering.  From these, it is clear that the dynamic DL-LES models generally better predict the resolved kinetic energy decay than the SM-LES calculations. The NM-LES is insufficiently dissipative, as could be anticipated.

The subsequent Figures \ref{Kinetic1024_7x}--\ref{Kinetic1024_15x_LogLog} illustrate the effect of increasing the filter size to $d=7$, $d=9$, and $d=15$ for the $N=1024^3$ case. For the larger $d=9$ and $d=15$ filter sizes, the embedded simulation is not DNS but rather a higher-fidelity NM-LES with $\Delta_\mathrm{Embedded}/\Delta_\DNSsub=3$ filter size. Overall, the dynamic DL-LES models produce the most-accurate predictions of the turbulent kinetic energy decay, even for relatively large filter sizes for which the embedded NM-LES does not resolve the Kolmogorov scale. The accuracy differences increase as the filter size increases, with both DL-LES models significantly outperforming the SM-LES and NM-LES benchmarks comparison models for the largest filter size.

\subsubsection{Resolved Kinetic Energy Spectra}

\begin{figure}[!htb]
   \begin{minipage}{0.48\textwidth}
     \centering
     \includegraphics[width=.9\linewidth]{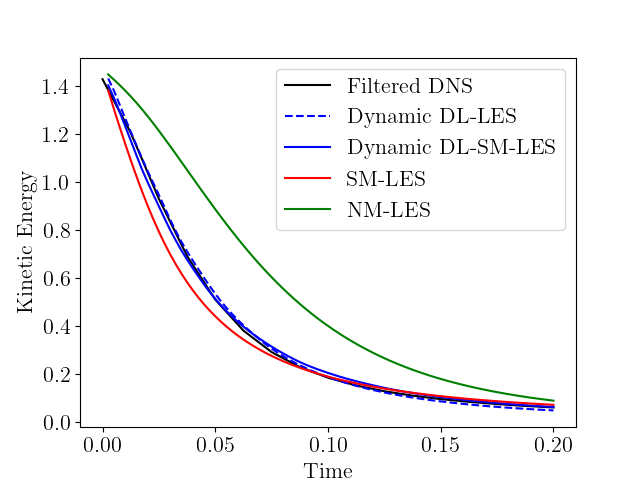}
     \caption{$512^3$ with $d=5$ filter}\label{Kinetic512}
   \end{minipage}\hfill
      \begin{minipage}{0.48\textwidth}
     \centering
     \includegraphics[width=.9\linewidth]{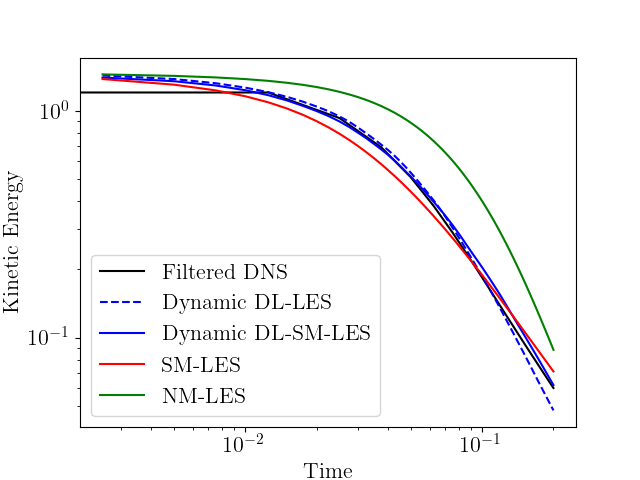}
     \caption{$512^3$ with $d=5$ filter}\label{Kinetic512_LogLog}
   \end{minipage}\hfill
   \begin{minipage}{0.48\textwidth}
     \centering
     \includegraphics[width=.9\linewidth]{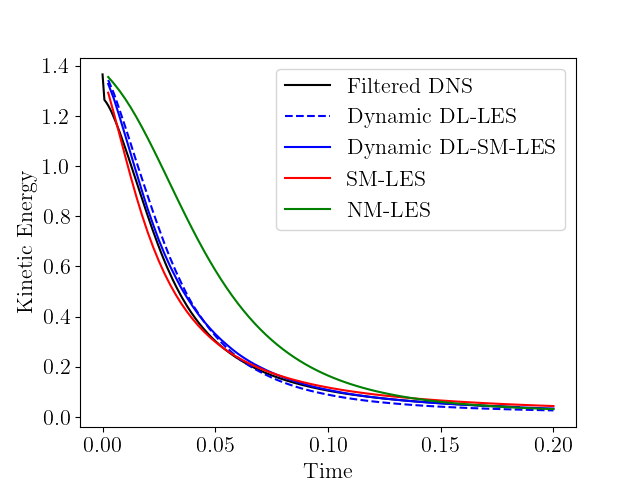}
     \caption{$1024^3$ with $d=5$ filter}\label{Kinetic1024_5x}
   \end{minipage}
      \begin{minipage}{0.48\textwidth}
     \centering
     \includegraphics[width=.9\linewidth]{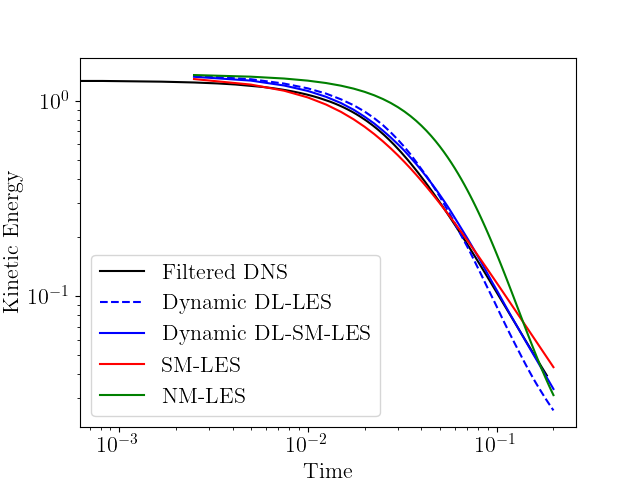}
     \caption{$1024^3$ with $d=5$ filter}\label{Kinetic1024_5x_LogLog}
   \end{minipage}
   
\end{figure}

\begin{figure}[!htb]
   \begin{minipage}{0.48\textwidth}
     \centering
     \includegraphics[width=.9\linewidth]{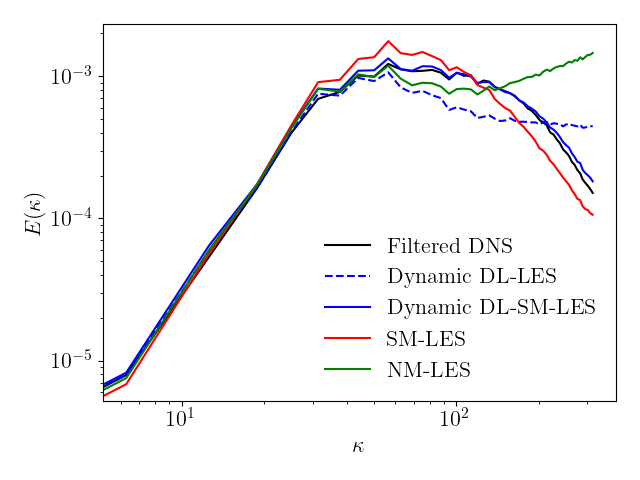}
     \caption{$512^3$ with $d=5$ filter}\label{Spectrum512_5x}
   \end{minipage}\hfill
   \begin{minipage}{0.48\textwidth}
     \centering
     \includegraphics[width=.9\linewidth]{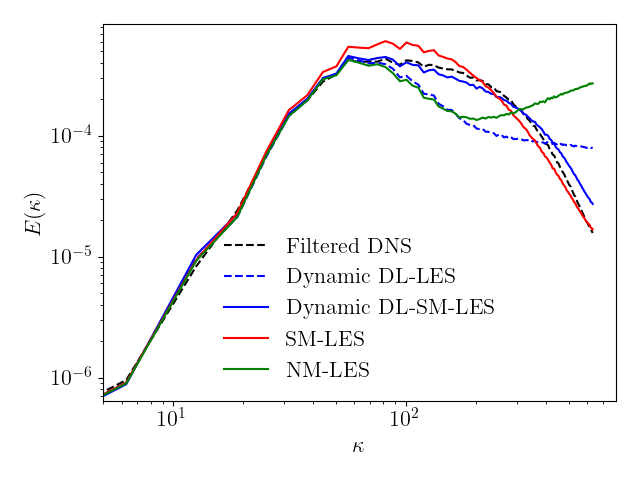}
     \caption{$1024^3$ with $d=5$ filter}\label{Spectrum1024_5x}
   \end{minipage}
   \begin{minipage}{0.48\textwidth}
     \centering
     \includegraphics[width=.9\linewidth]{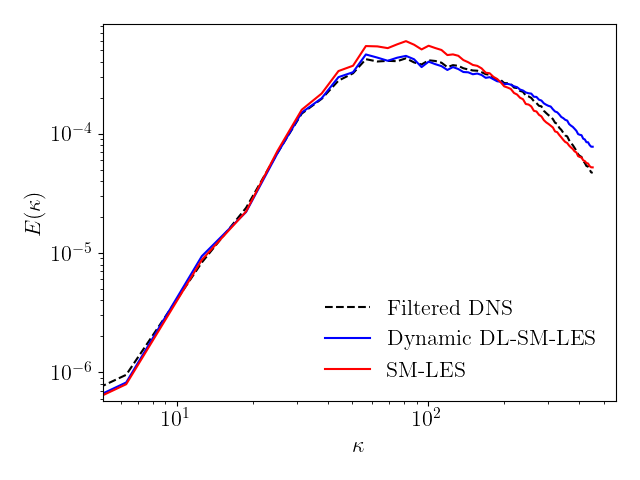}
     \caption{$1024^3$ with $d=7$ filter}\label{Spectrum1024_7x}
   \end{minipage}\hfill
   \begin{minipage}{0.48\textwidth}
     \centering
     \includegraphics[width=.9\linewidth]{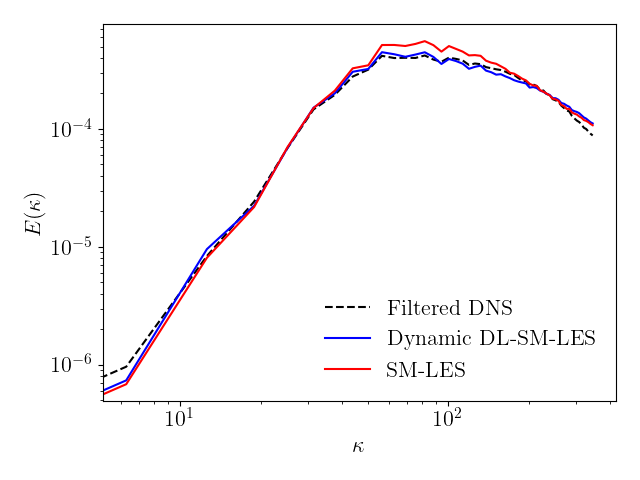}
     \caption{$1024^3$ with $d=9$ filter}\label{Spectrum1024_9x}
   \end{minipage}
      \begin{minipage}{0.48\textwidth}
     \centering
     \includegraphics[width=.9\linewidth]{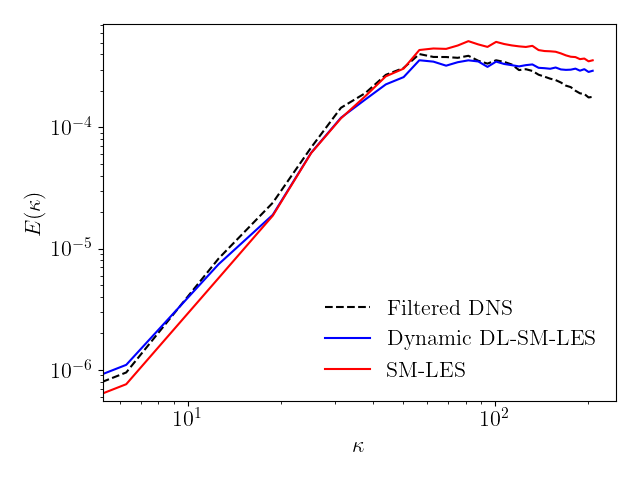}
     \caption{$1024^3$ with $d=15$ filter}\label{Spectrum1024_15x}
   \end{minipage}
   
\end{figure}

Figures \ref{Spectrum512_5x}--\ref{Spectrum1024_15x} compare the resolved kinetic energy spectra of the filtered DNS and LES models. Although the dynamic DL-LES models are not perfect, 
when coupled with the constant-coefficient Smagorinsky model (dynamic DL-SM-LES), 
they result in much more accurately predicted resolved kinetic energy at the highest wavenumbers compared to the Smagorinsky model alone. Conversely, the SM-LES often overestimates the resolved kinetic energy at the intermediate wavenumbers and, for the smaller filter sizes, underestimates the resolved kinetic energy at the highest wavenumbers. 
The NM-LES overestimates the resolved kinetic energy at the highest wavenumbers, which is physically incorrect but consistent with the lack of subgrid dissipation. Interestingly, the dynamic DL-LES model without the Smagorinsky term also overpredicts the energy at the highest wavenumbers, though not to the same extent as NM-LES, despite the fact that this DL-LES model accurately predicts the overall resolved kinetic energy decay (Figures~\ref{Kinetic512}--\ref{Kinetic1024_5x_LogLog}).

It should be noted that the resolved kinetic energy of the filtered embedded DNS from a coupled embedded DNS-LES simulation without a closure model simulation is as accurate as the dynamic DL-LES for predicting the resolved kinetic energy. Therefore, these results should be viewed as a proof-of-concept that the dynamic DL-LES can learn a correct closure model from the embedded DNS and generalize it from the embedded DNS sub-domain to the entire domain. Future research, discussed in more detail in Section \ref{Conclusion}, will evaluate the performance of the dynamic DL-LES method on inhomogeneous flows.

\section{Conclusion} \label{Conclusion}
We have developed a new dynamic closure model for large-eddy simulation. A deep learning closure model is dynamically learned from embedded DNS data using online optimization. Unlike previous machine learning closure models, high-fidelity training data is not required. This is a key advantage since generating DNS data can be computationally expensive or even intractable. Furthermore, the deep learning closure model is trained on data from the exact geometry/physical regime for which a prediction is being made. In previous machine learning methods, the closure model would be trained on data for a different geometry/physical regime. In future and ongoing research, we are working on evaluating the dynamic deep learning closure method for inhomogeneous flows such as spatially-evolving jet flows and turbulent flows around bluff bodies. In these flows, we will evaluate whether the method can accurately predict the steady-state statistics of the flows.

\bibliographystyle{plainnat}
\renewcommand{\refname}{References}

\end{document}